\documentstyle[epsfig,12pt]{article}
\newcommand{\p}{\mbox{e$^+$}}
\newcommand{\n}{\mbox{e$^-$}}
\newcommand{\pn}{\mbox{e$^+$e$^-$}}
\newcommand{\UTa}{\mbox{$^{238}$U + $^{181}$Ta\ }}
\newcommand{\UU}{\mbox{$^{238}$U + $^{238}$U\ }}
\newcommand{\UPba}{\mbox{$^{238}$U + $^{208}$Pb\ }}
\newcommand{\UPb}{\mbox{$^{238}$U + $^{206}$Pb\ }}
\newcommand{\Pba}{\mbox{$^{206}$Pb\ }}
\newcommand{\Pbb}{\mbox{$^{207}$Pb\ }}
\begin{document}
\begin{titlepage}
\vspace*{-0.7cm}
\begin{center}
{\Large \bf \mbox{Weak \pn {} lines from internal pair conversion}}

\vspace*{3mm}
{\Large \bf \mbox{observed in collisions of $^{238}$U with heavy nuclei}} 

\end{center}
 
\vspace*{7.0mm}
\begin{center}
{\large
 S.~Heinz$^a$, E.~Berdermann$^b$, F.~Heine$^a$,   
 O.~Joeres$^a$, P.~Kienle$^a$, I.~Koenig$^b$, 
W.~Koenig$^b$,
 C.~Kozhuharov$^b$, U.~Leinberger$^b$, 
 M.~Rhein$^b$, A.~Schr\"oter$^b$, H.~Tsertos$^{c}$
}

\vspace{0.4cm}
{\bf (The ORANGE Collaboration at GSI)}

\vspace{0.2cm}
$^a$ Technische Universit\"at M\"unchen,  D--85748 Garching, 
Germany\\
$^b$ Gesellschaft f\"ur Schwerionenforschung (GSI),
                                              D--64291 
Darmstadt, Germany\\
$^c$ University of Cyprus, CY--1678 Nicosia, Cyprus\\
\end{center}

\vspace*{0.2cm}
\begin{center}
\section*{Abstract}
 \end{center}
We present the results of a Doppler-shift correction to 
the measured \pn--sum-energy spectra obtained  
from \pn--coincidence measurements in \UPb and \UTa{} col\-li\-sions 
at beam energies close to the
Coulomb barrier, 
using an improved experimental 
setup at the double-Orange spectrometer of GSI.
Internal-Pair-Conversion (IPC) \pn{} pairs from discrete 
nuclear transitions
of a moving emitter have been observed following 
Coulomb excitation of the 1.844 MeV (E1) transition in 
\Pba  and neutron transfer to
the 1.770 MeV (M1) transition in \Pbb.
In the collision system \UTa,  IPC transitions were observed from the Ta-like 
as well as from the U-like nuclei. In all 
systems the Doppler-shift corrected \pn--sum-energy spectra show 
weak lines at the energies expected from the corresponding
$\gamma$--ray spectra with cross sections being consistent 
with the measured
excitation cross sections of the $\gamma$ lines 
and the theoretically predicted IPC coefficients. 
No other than IPC \pn--sum-energy lines were found in the 
measured spectra.
The transfer cross sections show a strong 
dependence on the 
distance of closest approach (R$_{min}$), thus signaling 
also a strong 
dependence on the bombarding energy close to the Coulomb 
barrier.\\

{\bf PACS numbers:} 14.60.Cd; 23.20.Ra; 25.70.Bc; 25.70.De; 25.70.Hi; 29.30.Aj
 
\end{titlepage}

\section{Introduction}

Previous results of the EPOS and ORANGE collaborations
at the UNILAC accelerator of GSI have revealed 
unexpected lines in the \pn--sum-energy spectra obtained 
in heavy-ion collisions at bombarding energies near the 
Coulomb barrier~\cite{Sal90,Ikoe93}.
No viable explanation could be found for these 
experimental results. 
At the beginning it was tempting to interpret these 
lines as being
due to the \pn{} decay of a previously unknown neutral 
particle with a mass
around 1.8 MeV/c$^2$, a conjecture which was ruled out 
by subsequent conclusive Bhabha-scattering experiments~\cite{Tse91}.   
 After this time it became clear that Internal Pair 
Conversion (IPC)
from excited nuclear transitions, which can in principle 
lead to narrow
lines in the \pn--sum-energy spectra, was not
sufficiently investigated and also theoretically
less well understood, such that some of the reported 
weak lines might be due to this process.
Only very recently, combined experimental and theoretical
efforts have provided a better understanding
of this fundamental decay channel of excited nuclear transitions 
in heavy nuclei at rest~\cite{Lein97}, and the method of how to 
observe weak lines from a moving emitter was developed 
systematically.  

The motivation of the present work was indeed a search 
for discrete 
IPC transitions, excited in heavy-ion collisions at 
energies close to the
Coulomb barrier, as possible candidates for previously 
observed weak
\pn-sum-energy lines with cross sections of the order of 
a few $\mu$b for the
\UTa  system and some tenth of $\mu$b for $^{238}$U + $^{208}$Pb 
collisions~\cite{Ikoe93}. 
Particularly for the last system, two
\pn--sum-energy lines at (575 $\pm$ 6) keV and at (787 
$\pm$ 8) keV 
have been observed at a beam energy of 5.9 
MeV/u~\cite{Ikoe93}. 
The 787 keV line
appeared in both quasielastic central (R$_{min} <$ 20 
fm) and 
rather peripheral (R$_{min} =$ 20--26 fm) collisions, 
whereas
the 575 keV line occurred only in coincidence with 
central collisions.
Both lines were seen without a Doppler-shift correction 
and only
in the opening-angle region $\theta_{\pn}$= 
$155^{\circ} - 177^{\circ}$, for which Doppler 
broadening is expected to be
small for similar energies of both leptons. 
The underlying \pn--energy-difference range was $-$100 keV $\leq$
$\Delta E = E_{e^+} - E_{e^-} \leq$ +34 keV, which also reflects 
the sweeping procedure of the Orange spectrometers. 
It is centered around
$\Delta E =$ $-$40 keV for the above observations. 

In order to test the response and sensitivity
of our experimental setup to IPC pairs from a moving 
emitter and to learn the observation methods for weak
lines we investigated first
a known transition in $^{206}$Pb populated via Coulomb 
excitation, using a 
$^{238}$U beam and a $^{206}$Pb target. 
On the other hand, since the number of combined nuclear 
charge (Z$_u=$174)
is the same for the collision systems \UPb and \UPba, 
the \pn{} lines seen
previously in the latter system~\cite{Ikoe93} should also 
appear
in the present experiment, if their origin is closely 
connected with 
the strong electromagnetic fields available in the 
collision. 
In addition, the \UTa collision system was investigated
with the objective to search for IPC sum-energy lines.

Here we report the first results from these improved 
investigations, being
performed at the UNILAC accelerator of GSI.
A dedicated Doppler-shift technique is exploited 
at the double-Orange setup~\cite{Lein96a,Shei96,Joe96}, 
which allows to reveal narrow lines in the corrected
\pn--sum-energy and $\gamma$-ray spectra from a moving 
emitter in an environment being dominated by continuous
spectral distributions. In addition 180$^{\circ}$
observation scenarios and data analysis were also exploited
for the observation of narrow \pn--lines.

It should be underlined here that the investigations 
and data analysis of this work  
are more far reaching with respect to Internal Pair Conversion
as possible source of \pn--lines than the experimental results 
reported most recently by the APEX 
collaboration~\cite{APEX95} at ANL and 
the EPOS II~\cite{Gan96} and ORANGE~\cite{Lein96b} 
collaborations at 
GSI, who have presented \pn--sum-energy spectra without a Doppler-shift 
correction, and
have all failed to observe previously reported narrow lines.

\vspace{0.5 cm}
\section{Experimental setup}

As shown in Fig. 1, leptons
emitted from a target placed between two toroidal 
magnetic field
spectrometers with their axis parallel to the beam 
direction were momentum
analyzed by the toroidal magnetic field and detected 
with two arrays 
of high-resolution Si detectors.
The toroidal $(\frac{1}{r}$)-field is generated by 60 
iron-free coils. 
Electrons with emission angles $\theta_{\n} = 38^{\circ} 
- 70^{\circ}$ and positrons with $\theta_{\p} = 
110^{\circ} - 145^{\circ}$ 
relative to the beam axis are accepted by the 
spectrometers. 
The lepton detectors consist each of 72 Si PIN diodes 
(chips)
of trapezoidal shape (base: 24 mm, top: 16 mm, height: 
16 mm, thickness: 1 mm) 
arranged in a Pagoda-like form~\cite{Lein96a}. 
Each chip is subdivided into three segments.
A Pagoda roof consists of six PIN diodes, i.e. 18 
segments. One detector array
is composed of 12 such roofs. 
Thus, we get a position sensitive detector with
216 segments read out in a matrix mode. At a given field 
setting only 
particles with a certain sign of charge are
focused onto the corresponding detector arrays. 
Thus, a very clean separation
of electrons and positrons is achieved. 

For further lepton identification the 
lepton energy and momentum is determined simultaneously. 
The 
deposited energy is measured by the PIN diodes, and the 
lepton momentum is calculated from the deflection and the
field setting. 
Only events for which the energy-momentum relation is fulfilled
are accepted, with the result that the remaining e$^+$ 
misidentification is
small and can be determined reliably. 
This is demonstrated in Fig. 2 for the case of the 
positron identification.
As can be seen a clear signature for the focussed 
positrons is achieved. 
In such a spectrum background processes like scattered electrons 
and $\gamma$ rays as well as
positrons backscattered from the detectors result in 
a broad continuous distribution which can be determined 
quantitatively.
In the case of the electron identification due to the 
higher electron
production probabilities the corresponding spectrum is 
practically 
background-free (see also Ref.~\cite{Lein96a}).

The spectrometer accepts lepton pairs with opening 
angles, $\theta_{e^+e^-}$, from  40$^{\circ}$ to 
$180^{\circ}$ in the laboratory system. 
This range 
can be subdivided into 10 bins of width $\approx \pm 
10^{\circ}$. 
For reconstruction of the reaction kinematics both 
scattered heavy ions are 
detected by 19 Parallel Plate Avalanche Counters (PPACs) 
which accept ions
scattered under polar angles of $\theta_{ion} = 
12.5^{\circ} - 35^{\circ}$ 
and $\theta_{ion} = 40^{\circ} - 70^{\circ}$ with a 
resolution of 
$1.0^{\circ}$ and $0.5^{\circ}$, respectively. 
The azimuthal-angle resolution of all (ion and lepton) 
detectors is 
$\Delta\phi=20^{\circ}$. \\
For the detection of $\gamma$ rays we used a high 
resolution
70\%-Ge(i) detector placed
at $\theta_{\gamma} = 86^{\circ}$ relative to the beam 
axis at a 
distance of 40 cm from the target.
An ionization chamber installed at $\theta = 40^{\circ}$ 
(not shown in Fig. 1) measures 
the energy of scattered particles and thus controls the 
effective target
thickness. It is also used for current normalization.  

Extensive measurements carried out with
radioactive $^{90}$Sr and $^{207}$Bi sources proved that 
our setup is
capable of detecting IPC pairs and of determining their 
opening-angle
distribution from an emitter at rest~\cite{Lein97,Lein96a}.
The measured FWHM of these
sum-energy lines is $\sim$16 keV, consistent with the 
sum-energy resolution 
of the lepton detectors.

\vspace{0.5 cm}
\section{Doppler-shift correction}

In the case that the leptons are emitted from a moving 
source, i.e. from an 
outgoing ion which moves with a reduced velocity 
$\beta_{ion}=(\upsilon_{ion}/c$),
their energy measured in the laboratory is affected by a 
Doppler shift.
The kinetic energy of the leptons in the rest frame of 
the ion $E_{cm}$ and 
their corresponding energy $E_{lab}$ in the laboratory 
system are connected 
by a Lorentz transformation   

\begin{equation}
   E_{cm} = \gamma_{ion}\left(E_{tot} - \sqrt{E_{tot}^2 
- m_{0}^2c^4}
  \times\beta_{ion}\cos\alpha\right) - m_0c^2,
\end{equation}

where $E_{tot}=E_{lab} + m_0c^2$, and $\alpha$ being the 
relative angle 
between the lepton and the emitting ion which depends on 
their polar and azimuthal angles as follows
\footnote{The angles refer to the laboratory system,
unless explicitly stated otherwise}.

\begin{equation}
\cos\alpha = 
\sin\theta_{lepton}\sin\theta_{ion}\cos(\phi_{ion}-\phi_
{lepton})
                + \cos\theta_{lepton}\cos\theta_{ion}.
\end{equation}
   
The reduced velocities of the outgoing ions after the 
collision are given
below: 

\begin{equation}               
\beta_{ion}^p = \frac{\sqrt{A_p^2 + A_t^2 + 2A_pA_t 
\cos\theta_{cm}}} 
               {\left( A_p + A_t \right)}\beta_p
\end{equation}

\begin{equation}
\beta_{ion}^t= \frac{2A_p\sin(\theta_{cm}/2)} 
         {\left(A_p + A_t\right)} \beta_p,
\end{equation}

where $A_p$ and $A_t$ are the mass numbers of the 
projectile and 
target ion, respectively, $\theta_{cm}$ is the heavy-ion 
scattering angle 
in the c.m. system, and \,$\beta_p = 
\sqrt{2(E_p/A_p)/931.5}$\,\, 
is the reduced velocity of the incident ion,
with $(E_p/A_p)$
being the bombarding energy per nucleon in units 
[MeV/u].

From the above relations it is obvious that the correction 
of the energy (Doppler shift)
(Eq. (1)) and its uncertainty (Doppler broadening) 
depends mainly on the
experimental polar and azimuthal angular determination  
and resolution of the detected leptons and
heavy ions. 
In our setup, the dominant contribution to the Doppler 
broadening is due to
the lepton polar angles which cannot be determined 
within the angular range
accepted by the spectrometers, and only weighted mean
values of $\overline{\theta}_{lepton}=$55$^\circ$ and 
$\overline{\theta}_{lepton}=$125$^\circ$ are
used, for the forward and backward spectrometer, 
respectively.    

The response and sensitivity of our setup to reveal 
narrow lines emitted from a
moving ion after applying an event-by-event 
Doppler-shift correction to the
measured energies has been carefully investigated by 
Monte Carlo simulations.
In such calculations lepton pairs from an IPC transition 
are generated in the
rest frame of the emitting ion with energy distributions 
taken from theoretical
calculations~\cite{Hof} and assuming an
isotropic \pn--opening-angle distribution. 
The energies of the leptons are transformed into the
laboratory system using the inverse transformation of 
Eq. (1).
The simulated events are then analyzed with the same 
analysis program used to
analyze the experimental data.
In a second step, the laboratory energies of the leptons 
are corrected                  
event by event using Eq. (1) and taking into account the 
angular resolution of our setup by a Monte Carlo 
procedure.

An illustrative example of such calculations is shown in 
Fig. 3 for the
822 keV IPC (E1) transition in $^{206}$Pb, populated by 
Coulomb excitation in
\UPb collisions at a beam energy of 5.93 MeV/u. 
The solid broad distribution
shows the \pn--sum-energy spectrum expected in the 
laboratory system if an emitter velocity of 
$\upsilon_{ion} \approx$ 0.04c $-$ 0.08c
is considered, as it is the case for peripheral collisions
with $\theta_{ion,cm} = 50^{\circ} - 70^{\circ}$.
It is
centered around 815 keV and has a FWHM of about 90 keV. 
The dashed narrow
peak represents the same events after an event-by-event 
Doppler-shift
correction is applied to the laboratory energies. The 
peak is centered at the
expected position ($\sim$822 keV) and has a FWHM of 
$\sim$16 keV. 
The latter is mainly due to the uncertainty of the 
lepton polar 
angles, discussed above. 
For central collisions, a velocity
of $\upsilon_{ion} \approx$ 0.1c for the recoiling ion is expected which
leads to sum-energy distributions with widths of 
about 100 keV in the laboratory system and of
nearly 40 keV after Doppler-shift correction. 

It should be pointed out here that the detection of IPC 
\pn--sum-energy lines
in the real experiment is rather difficult because they 
are produced 
with weak intensities, superimposed on a 
continuous spectral distribution (see below)
mainly caused by 
pair creation in the time-changing high Coulomb field
of the high-Z collision partners (dynamic \pn{} creation). 
It is particularly difficult to detect IPC lines
from projectile-like nuclei in grazing collisions, because the
Doppler shifts and their resulting corrections are rather large. 
In such cases
the sensitivity can be improved by restricting the analysis to
events with \pn--opening angles of about $180^{\circ}$ or by
eliminating the most peripheral scattering events 
with R$_{min} \geq$ 25 fm.

\vspace{0.5 cm}
\section{Experiment and results}

\subsection{The collision system \UPb}

Data were taken for the collision system $^{238}$U + 
$^{206}$Pb using
$^{238}$U beams and 800 $\mu$g/cm$^2$ thick $^{206}$Pb 
targets mounted on a rotating target wheel. 
The projectile energy (5.93 MeV/u) is slightly below the 
Coulomb 
barrier (6.06 MeV/u). The $3^-$-level at 2.65 MeV
in $^{206}$Pb is populated via Coulomb excitation and 
deexcites
for the most part into the lower lying $2^+$-level with 
a 
$\gamma$-transition energy of \mbox{1844 
keV}~\cite{Led78}.
Following an event-by-event Doppler-shift correction to 
the Pb-like 
recoiling ion, a pronounced $\gamma$ line with a total 
excitation 
cross section of $\sigma_{\gamma}=$\mbox{(55 $\pm$ 5) 
mb} appears at 
(1844 $\pm$ 1) keV in the measured energy spectrum (Fig. 4a). 
Its FWHM is 22 keV, mainly caused by uncertainties 
in the Doppler-shift correction introduced by the 
finite angular resolution of the heavy-ion detectors.

The excitation probability, P$_{\gamma}$(R$_{min}$), was 
determined as
a function of distance of closest approach, R$_{min}$, 
by normalizing the
$\gamma$ yield in certain R$_{min}$ intervals with the 
corresponding number
of elastically scattered ions. R$_{min}$ was determined from
the measured scattering angle $\theta_{ion,cm}$ assuming
Rutherford trajectories. 
At large R$_{min}$ values, P$_{\gamma}$ shows an
exponential decrease, whereas at small R$_{min}$ values 
P$_{\gamma}$ is cut off
abruptly at R$_{min} \sim$ 17 fm, where the nuclei come 
into contact.
Such a behaviour is typical for Coulomb excitation (Fig. 4b).
The results are in accordance with measurements of the 
EPOS collaboration who 
investigated the same collision system at a bombarding 
energy of 5.82 MeV/u and
$\sim$400 $\mu$g/cm$^2$ thick targets. They reported 
recently a measured 
cross section of
$\sigma_{\gamma}=$ (44 $\pm$ 7) mb~\cite{Bau96}.

From the 1844 keV transition we expect IPC pairs with a 
sum energy of 
\mbox{822 keV.} Multiplying the measured $\gamma$ cross 
section 
with the theoretically predicted IPC coefficient of 
$\beta = 4.0 \times 10^{-4}$, for an 
electromagnetic
transition with multipolarity E1, Z=82 and 1850 keV
energy~\cite{Schl79}, we
expect a total cross section for IPC of 
$\sigma_{IPC}=$(22 $\pm$ 2) $\mu$b. 
In order to optimize
the peak-to-continuum ratio in the e$^+$e$^-$-sum-energy 
spectra, we eliminated large positive energy differences 
$\Delta$E = E$_{e^+} -$ E$_{e^-} >$ 175 keV, 
and accepted only rather peripheral collisions with 
R$_{min} \geq$ 23 fm. 
Thus, we could reduce the contribution of the continuum 
pairs produced by the large time-changing Coulomb field 
and combinatorial background from coincidences of 
$\delta$ electrons emitted in the same collision with positrons.
Particularly, negative energy differences (E$_{e^+} -$ E$_{e^-}$) select
the combination of low e$^+$ energies and high e$^-$ energies, thus
reducing significantly the contribution of the exponentially increasing 
$\delta$-electron rate at low energies.
The event-by-event Doppler-shift corrected 
\pn--sum-energy spectrum 
obtained under these conditions is shown in Fig. 4c. 
The spectrum is integrated over the whole range of 
lepton opening angles covered experimentally.
As shown, the continuous part of the measured spectrum is well 
reproduced by a reference distribution (smooth solid curve), 
being gained by an event-mixing procedure. 

From the IPC cross section quoted above, a total of 
about 35 IPC pairs is 
then expected in the spectrum shown in Fig. 4c, using an 
IPC detection 
efficiency 
of $\epsilon_{IPC} =$ (1.6 $\pm$ 0.2) $\times 10^{-3}$.
They should result in a line at an energy of $\sim$820 
keV with a FWHM 
of $\sim$ 20 keV, superimposed on the \pn{} continuum.
The IPC detection efficiency has been obtained by a 
Monte Carlo simulation, which assumes
isotropically emitted e$^+$e$^-$ pairs and 
theoretically-calculated 
lepton energy distributions for Z $=$ 82 and E1 
multipolarity~\cite{Hof}. 
It should be noted here that in this case the assumption of isotropically 
emitted \pn{} pairs is a good approximation, as shown by the 
analysis of the measured IPC line.
The sweep for these
measurements was performed such that lepton pairs with
rather positive energy differences E$_{e^+} -$ E$_{e^-}
\approx$ 200 keV were detected with a maximum efficiency.
The total IPC pair detection efficiency was (2.9 $\pm$ 0.4) $\times 10^{-3}$
for this sweep mode. The elimination of pairs with energy 
differences $\Delta$E $>$ 175 keV leads to the reduced
detection efficiency quoted above. 
Note also here that previous measurements were differently 
performed, such that pairs with E$_{e^+} -$ E$_{e^-}
\approx$ $-$40 keV were detected with the maximum efficiency, resulting
in a total detection efficiency of 3 $\times 10^{-3}$ and a
better signal to continuum ratio for the detection of IPC
lines.
The width of the Doppler-shift corrected line depends
on the emitter velocity and on the angular resolution of 
the detectors. 
As pointed out in sect.~3, for the case of a rather slow 
emitter (i.e., $\upsilon_{ion} \approx$ 0.05c), as obtained from peripheral
collisions and looking at the target-like excited nucleus, 
we expect line widths near the limit given by the detector 
sum-energy resolution, i.e., FWHM $\sim$ 16 keV~\cite{Lein96a}.
Figure 4c shows an excess, relative to the
continuous distribution determined by event mixing,
of (40 $\pm$ 12) counts in the energy window 805 keV to 825 keV
and (57 $\pm$ 14) counts in the energy window between 805 keV
and 835 keV.
This corresponds to differential
cross sections of (d$\sigma$/d$\Omega_{cm}) =$ (3.1 $\pm$ 1.0) $\mu$b/sr
and (d$\sigma$/d$\Omega_{cm}) =$ (4.4 $\pm$ 1.1) $\mu$b/sr, respectively,
for $\theta_{ion,cm} = 40^{\circ} - 70^{\circ}$ (R$_{min} = $ 23 $-$32 fm).

Figure 5a shows the uncorrected spectrum for the whole opening
angle range of observation ($\theta_{\pn} = 40^{\circ} - 180^{\circ}$). 
The spectrum was
obtained under identical conditions to that shown in Fig. 4c. 
Without a Doppler-shift
correction, no line is apparent at an energy around 820 keV and only a
slight surplus can be recognized. 
According to a Monte Carlo simulation, we
expect the IPC events from this transition in a broad 
distribution of
FWHM $\approx$ 80 keV for R$_{min} = $ 23 $-$ 32 fm, thus leading to a 
undetectable signal above the continuum.

As an important result with respect to previous observations we note that 
the weak 820 keV line is also present in the  
uncorrected \pn--sum-energy spectrum, shown in Fig. 5b, when the \pn--opening 
angles are restricted to the range
$140^{\circ} \leq \theta_{\pn} \leq 180^{\circ}$.
For this particular case of $\theta_{\pn}$ selection, the leptons are emitted 
nearly back to back with the result that their Doppler shifts 
cancel for leptons with similar energies. 
What we find in Fig. 5b, which was gained under the same kinematical
conditions than Fig. 4c, is a surplus of (14 $\pm$ 7) events 
in the energy window 805 keV to 825 keV and (23 $\pm$ 8) events
in the energy window between 805 keV and 835 keV. This
corresponds to differential cross sections of (d$\sigma$/d$\Omega_{cm}) =$
(1.1 $\pm$ 0.6) $\mu$b/sr and (1.8 $\pm$ 0.6) $\mu$b/sr, respectively,
for $\theta_{ion,cm} = 40^{\circ} - 70^{\circ}$ (R$_{min} =$ 23 $-$ 32 fm).
The Monte Carlo simulation of the expected line shape for the back-to-back 
scenario following the E1 transition is presented in Fig. 5c.   
The kinematical conditions are identical to those of Fig. 4c 
(i.e., R$_{min} =$ 23 $-$ 32 fm, and $\Delta$E $\leq$ 175 keV).
According to the simulations, we would expect in this spectrum about 
12 IPC events at a position of about 820 keV. 

We found an additional line in the $\gamma$ spectra, 
when corrected on 
the Pb-like scattered ion, at an energy of (1770 $\pm$ 1) keV.
This line appears only in central collisions with 
R$_{min} <$ 20 fm (Fig. 6a). 
In the R$_{min}$ parameter range selected
(17 fm $<$ R$_{min}$ $<$ 20 fm), the $\gamma$ lines at 
1770 keV and at 
1844 keV have comparable intensities. 
The excitation probability for the 1770 keV line
as a function of R$_{min}$, is shown in Fig. 6b. 
It peaks within a narrow R$_{min}$ interval, typical for 
transfer reactions, 
for which the transfer probability is expected to become 
large when the 
nuclei come into contact. 
This also means 
that the excitation function should exhibit a rather 
narrow structure at
energies close to the Coulomb barrier with a width of 
about 0.3 MeV/u. 
Similar narrow structures in the R$_{min}$ dependence 
were observed for a
two-neutron transfer reaction, leading to the known 3.71 
MeV $5^-$ level 
in $^{208}$Pb~\cite{Led78}.

The line at 1770 keV is assigned to a known M1 
transition in $^{207}$Pb
from the $\frac{7^-}{2}$ (2.34 MeV) to $\frac{5^-}{2}$ 
(0.57 MeV) 
state~\cite{Led78}, populated by neutron transfer 
from $^{238}$U to $^{206}$Pb, with a total
cross section of $\sigma_{\gamma}=$(1.1 $\pm$ 0.3) mb. 
The corresponding IPC production cross
section is expected to be 
$\sigma_{IPC}=$(0.3 $\pm$ 0.1) $\mu$b with 
$\beta = 2.8 \times 10^{-4}$~\cite{Schl79}, which should 
lead to a weak 
\pn-sum-energy line at $\sim$750 keV  
with a FWHM of $\sim$ 40 keV, and an intensity of about 
10 counts. In the corresponding Doppler-shift corrected
\pn--sum-energy spectrum, shown in Fig. 6c, no surplus
can be found in the energy window between 728 keV and 768 keV
and one cannot distinguish  
the appearance of the line from the expected statistical 
fluctuations, which are of the same order.
This leads to a cross section limit of $<$ 0.9 $\mu$b/sr. 
The situation is similar for the line at $\sim$820 keV, 
which is expected to have a comparable intensity.

Note that the 1770 keV $\gamma$ line was not observed by 
the EPOS 
collaboration, who investigated the same collision 
system 
at a beam energy of 5.82 MeV/u and
a $\sim$ 400 $\mu$g/cm$^2$ thick $^{206}$Pb 
target~\cite{Bau96} compared with a beam energy
of 5.93 MeV/u and a 800 $\mu$g/cm$^2$ thick
$^{206}$Pb target in our experiment.
This may also be taken as an indication for a strong
beam energy dependence of transfer reactions at the 
Coulomb barrier of heavy collision systems.

Figure 7a shows a $\gamma$-ray spectrum from the collision system
\UPb after applying an event-by-event Doppler-shift correction to the U-like
heavy ion scattered in the R$_{min}$ range 19.5--23 fm. 
The spectrum
reveals a line at (1778 $\pm$ 2) keV with a total excitation cross
section of $\sigma_{\gamma}=$ (6 $\pm$ 1) mb. 
The measured excitation probability
P$_{\gamma}$(R$_{min}$), shown in Fig. 7b, clearly points to a
Coulomb-excited transition in $^{238}$U. 
The line was also observed by the EPOS~\cite{Bau96,Baer95}
and APEX~\cite{APEX95b} collaborations as well as 
by Ditzel et al.~\cite{Dit96}. Recently an experiment was carried
out by Zilges et al.~\cite{Zil95} to search for strong dipole
excitations around 1.8 MeV in $^{238}$U. In this experiment the existence 
of the 1780 keV transition in $^{238}$U is confirmed and classified as
electric dipole (E1) transition. 
For estimating the number of IPC pairs to expect from the
1778 keV $\gamma$ line we assumed an E1 transition
for which the largest IPC coefficient is expected.
Calculating with an IPC coefficient of
$\beta = 3.3 \times 10^{-4}$, an upper limit for the total
cross section for IPC pair production of $\sigma_{IPC} \leq $2 $\mu$b 
is expected.

A main problem arises thereby with respect to the performance of the 
Doppler-shift technique which is common to all experiments 
using $^{238}$U beams, with the exception of \UU collisions. 
In peripheral
collisions, in which the continuum background is small, the velocity of
the $^{238}$U--like ions is rather high, thus leading to
IPC-line widths of nearly $\sim$40 keV in the Doppler-shift corrected
\pn--sum-energy spectra. 
This is mainly due to our
limited angular resolution for \p{} and \n{} emission angles, discussed in
sect. 3.
The broadening of the \pn{} lines worsens the  
peak-to-continuum ratio
in the corrected \pn--sum-energy spectra, thus making IPC lines of
small production yield unobservable. 
Therefore we tried a different approach.
Figure 7c shows the \pn--sum--energy spectrum, corrected for Doppler
shifts, assuming emission from  
the U-like ion scattered in the R$_{min}$ range 19.5--23 fm.
In this case the emitter velocity is rather low 
($\upsilon_{ion} \approx$ 0.05 c) and hence
the magnitude of the resulting Doppler shift is low, too. 
The corresponding \pn--energy 
difference ranges from $-$200 to 175 keV.

From the spectrum shown in Fig. 7c, a cross section limit 
of (d$\sigma$/d$\Omega_{cm}) \leq 0.25 \mu$b/sr for 
$\theta_{ion,cm} = 72^{\circ} - 100^{\circ}$ (R$_{min} =$ 19.5 $-$ 23 fm)
is derived, which has to be
compared with a calculated 
value of (d$\sigma$/d$\Omega_{cm}) \leq 0.17 \mu$b/sr.
The latter value has been obtained 
assuming an E1 transition which leads to an \pn--sum-energy line at 
$\sim$758 keV with a width of $\sim$25 keV.
Obviously, our experimental sensitivity has reached the detection limit 
in this case. 
In \UU collisions studied in earlier
experiments, the U-like recoil has a low velocity which allows easier
detection of IPC transitions, particularly in 180$^{\circ}$ geometry, 
without need of Doppler-shift corrections, and only negative energy differences
were accepted which increases the signal-to-background ratio for IPC
transitions. 

\subsection{The collision system \UTa}

With the collision system \UTa data were taken using a 
$^{238}$U beam at a bombarding energy of 6.3 MeV/u (slightly above the
Coulomb barrier) and 1000 
$\mu$g/cm$^2$ thick $^{181}$Ta targets. 
After an event-by-event Doppler-shift correction to the Ta-like recoiling 
ion, a number of $\gamma$ lines with cross sections between some
mb and some 10 mb and energies below 1600 keV appear in the
corrected $\gamma$--ray spectrum shown in Fig. 8a. The corresponding
transitions are obviously hitherto unknown in the $^{181}$Ta
nucleus~\cite{Led78}, but they were also measured with comparable cross 
sections by the EPOS~\cite{Baer95} and APEX~\cite{APEX95b} 
collaborations as well as by Ditzel et. al. \cite{Dit96}.

Figure 8b shows the excitation probability 
P$_{\gamma}$(R$_{min}$) as a function of R$_{min}$ for
the strongest $\gamma$ transition at E$_\gamma=$(1380 $\pm$ 2) keV, which is 
representative for all $\gamma$ lines observed
between 1000 and 1600 keV. 
The dependence of the excitation probability on R$_{min}$
is different than for known low energy $\gamma$ lines in
$^{181}$Ta from Coulomb excitation. 
One notes particularly the nearly
constant behaviour in the R$_{min}$ range between 20 and 25 fm. 
What is more surprising in this context is the fact that some of
the observed $\gamma$ lines have energies which could be
attributed within 1--2 keV to known transitions in
$^{184}$W \cite{Led78}. 
Therefore one can not completely rule out
a contribution from (1p2n) transfer reactions from
$^{238}$U to $^{181}$Ta at energies close to the Coulomb
barrier, governed by an anomalous deflection function
of the scattering process.

All electromagnetic transitions observed are 
accompanied by \pn--pair production via IPC
with expected total cross sections between
several 0.1 $\mu$b and several $\mu$b, assuming IPC
coefficients of the order of $10^{-4}$. 
In this case
we should be able to observe the corresponding \pn--sum-energy lines 
after appropriate Doppler-shift correction, as shown 
in the case of the \UPb collision system.
However, because the energies of the IPC lines from the
strongest $\gamma$ transitions are expected at \pn--sum energies 
below 500 keV, their detection efficiency is strongly
reduced by our particular choice of the momentum
acceptance of both spectrometers. 
The spectrometer momentum acceptance was optimized
for maximum detection efficiency at around 600 keV,
dropping off slowly towards higher energies and
relatively abruptly to lower energies, in order to 
search efficiently for lines  at \pn--sum energies around 600 keV
(see Refs.~\cite{Ikoe93,Lein96b}).
As a consequence of this efficiency cutoff
at low energies, we can only observe in our experiment 
the high-energy
part of the \pn{} IPC pairs belonging to the energetically
broad $\gamma$ distribution between 1500 and 1580 keV
with a total cross section of 
$\sigma_\gamma=$(15 $\pm$ 3) mb. 
In fact
we can only expect to detect this part of the
IPC spectrum centered around 550 keV which should lead
to a rather narrow structure in the Doppler-shift corrected
\pn--sum-energy spectrum with a width of about
30--35 keV resulting from a Monte Carlo simulation (Fig. 9b).
It is produced by several close-lying
transitions in the energy range centered around
1550 keV. Assuming an IPC coefficient of
2$\times$ $10^{-4}$ for a transition energy of 1550 keV
with E1 multipolarity in a nucleus with Z $=$ 73, a
total cross section for IPC pair production of
$\sigma_{IPC}=$(3.0 $\pm$ 0.8) $\mu$b is expected from the observed
$\gamma$ strength. 

Figure 9a shows the \pn--sum-energy spectrum in coincidence
with one ion, scattered into the angular range:
$40^{\circ} \leq \theta_{ion,cm} \leq 62^{\circ}$
(i.e., 22.4 fm $\leq$ R$_{min}$ $\leq$ 30 fm).
In contrast to collisions with quasielastic 
R$_{min}$-dependance, which are governed by two-body kinematics, 
the U-Ta collisions, leading to excited nuclear states above 1 MeV,
show a more complex R$_{min}$-dependance accompanied by a loss
of events by a factor of 2.5
in the two-body final channel. Therefore we show in Fig. 9a
only the \pn--spectra in coincidence with one 
scattered ion (Ta-like), in order to obtain better statistics.
In Fig. 9a the energies are corrected for 
Doppler shifts, assuming a Ta-like ion as the
emitter of the observed pairs. 
In order to suppress the
contribution of the continuum \pn--pairs from the strong
time-changing Coulomb field of both collision partners, 
very positive energy differences between the
positrons and electrons were cut out, and we accepted
only events which fall in an \pn--energy-difference
window between $-$150 and $+$100 keV. 
Furthermore, only
rather peripheral collisions with R$_{min} >$ 22 fm
are accepted. In order to reduce the magnitude of 
Doppler shifts to be corrected, the acceptance is
restricted to low velocity Ta-like recoils. This
ensures a narrow IPC line width and thus improves the
signal-to-background ratio. 
In the \pn--sum-energy
spectrum shown in Fig. 9a, a line structure with FWHM $=$ 30 keV at an
energy of (558 $\pm$ 10) keV with a surplus of (64 $\pm$ 21) counts 
in the energy range between 543 keV and 573 keV appears above
the continuum, which is constructed by event mixing. 
As already pointed out above, the line shape
is well reproduced by a Monte Carlo simulation shown in Fig. 9b.
On the
basis of the measured $\gamma$--ray spectrum, one would expect at an energy
of $\sim$560 keV a line structure of about 8 counts only, as expected 
by an IPC detection efficiency of 
$\epsilon_{IPC}=(0.75 \pm 0.05) \times 10^{-3}$. 
This value was obtained by
a Monte Carlo simulation assuming an isotropic 
\pn{} angular distribution and multipolarity E1.
The 64 counts observed correspond to a
differential cross section of 
(d$\sigma$/d$\Omega_{cm})=(18 \pm 6$) $\mu$b/sr 
in the scattering angle range $40^{\circ} \leq 
\theta_{ion,cm} \leq 62^{\circ}$, which is a
factor of 8 larger than expected from the corresponding
$\gamma$-ray yield. 
This discrepancy is surprising and not yet understood. 
A possible explanation for the high line intensity might be due to
an admixture of an E0 to an E2 transition between rotational states
of an excited vibrational band to a state with the same spin I $\geq$ 2 
of the ground state rotational band.

Very interesting to note here is the fact that the line at 558 keV 
appears also in
the uncorrected \pn--sum-energy spectrum shown in Fig. 10, where the
\pn--opening-angles are restricted to the range 
$140^{\circ} \leq \theta_{\pn} \leq 180^{\circ}$ and only 
energy differences $\leq$ 100 keV are
accepted. The corresponding R$_{min}$ range is 22.4 to 30 fm like in 
Fig. 9a. The line at (558 $\pm$ 10) with a FWHM of (20 $\pm$ 2) keV 
has an intensity of (31 $\pm$ 11) counts in the energy region between 548 keV
and 568 keV above the continuum. 
The line position and width are in good accordance with our 
expectations from Monte Carlo simulations, but the intensity of the line
is nearly a factor of 2 larger than expected from the experimental
results in Fig. 9a which is also not well understood.

From the collision system \UTa one should also expect IPC pairs from the
Coulomb-excited 1778 keV transition, discussed at the end of the
previous chapter. 
But two criteria make this system
less favourable for detecting IPC lines: the total cross section of the
excited 1778 keV $\gamma$ transition is with 2 mb a factor of 3 reduced
compared to the cross section obtained from the collision system
\UPb. 
Moreover, the \pn--pair-detection efficiency at sum-energies around 760 keV
is strongly reduced because of a different choice of the momentum
acceptance of our spectrometers as compared to the \UPb
measurements.

\vspace{0.5 cm}
\section{Summary and Conclusions}

In summary, we reported the observation of $\gamma$-transitions which
could contribute to observable IPC lines in collisions of $^{238}$U with
$^{206}$Pb and $^{181}$Ta nuclei at bombarding energies close to 
the respective Coulomb barrier. The $\gamma$ lines were observed after a
Doppler-shift correction of the spectra with cross sections of several
10 mb in the case of Coulomb excitation and in the order of 1 mb for
transfer reactions. We succeeded also to observe for most of the stronger
nuclear transitions weak IPC lines in Doppler-shift corrected 
\pn--sum energy spectra with the expected cross sections of typical
several $\mu$b, in accordance with theoretically predicted IPC
coefficients in the case that the transition multipolarity was known. 

The transitions from Coulomb excitation show an expected 
dependance on the minimum distance of closest
approach (R$_{min}$), with a sharp drop of the excitation
probabilities at smaller R$_{min}$ values, caused by nuclear
contact, and an exponential decrease for larger R$_{min}$ values.
The nuclear excitation probabilities following transfer
reactions at bombarding energies close to the Coulomb
barrier show a quite different R$_{min}$ dependance which
is sharply peaked around a typical interaction distance.
This reflects also a very strong threshold energy dependance
at the Coulomb barrier. The R$_{min}$ dependance of high
excitations observed in $^{238}$U $+$ $^{181}$Ta collisions is
less well understood. This may reflect a
not yet studied behaviour of multi-nucleon transfer
reactions at bombarding energies close to the barrier. 

Some weak \pn--sum-energy lines, previously observed in \UTa and 
$^{238}$U $+$ $^{208}$Pb collisions by the Orange 
collaboration~\cite{Ikoe93},
show characteristics of IPC lines observed in
our present work.
According to the present experience, they may have well been caused by
nuclear transitions accompanied by $\gamma$ transitions
observable only in high-resolution Doppler-shift corrected
$\gamma$-ray spectra, Doppler-shift corrected \pn--sum-energy
spectra and 180$^{\circ}$ sum-energy spectra, like those presented here. 
Unfortunately, for the previously investigated collision system 
$^{238}$U $+$ $^{208}$Pb at a bombarding energy of 5.9 MeV/u, for which we 
have observed at
selected $180^{\circ}$ \pn{} emission angles, \pn--sum-energy lines 
at 575 keV and 787 keV~\cite{Ikoe93}, the corresponding $\gamma$--ray spectra
were measured with a low energy resolution 
\mbox{Na(Tl)I} detector~\cite{Ikoe93} and thus the observation
of weak $\gamma$ lines was not possible.
Therefore these lines might have been accounted for
by IPC transitions because for a $180^{\circ}$
\pn{} observation geometry in the Orange
setup, the Doppler shifts nearly cancel in case of
comparable \p{} and \n{} energies as has been shown in Fig. 5b. 
Note also that the 575 keV and 787 keV \pn{}
lines, observed previously in $^{238}$U $+$ $^{208}$Pb collisions, were not
found in our present investigations of the collision system 
$^{238}$U $+$ $^{206}$Pb. 
This clearly means that
strong field effects can not be the origin of these lines. 

There is still the unresolved problem of 
$0^+ \rightarrow 0^+$ monopole transitions in 
\pn--pair spectroscopy. 
Until now no E0 transition in the interesting energy region could
be identified uniquely in $^{238}$U induced heavy-ion collisions.
Recently a search for $0^+ \rightarrow 0^+$ monopole
transitions in \UTa collisions at a beam energy of
6.0 MeV/u was conducted at GSI using conversion electron
spectroscopy~\cite{Dit96}. A limit of 0.3 mb for the production cross
section of a K-conversion line in $^{181}$Ta at an energy 
around 1.8 MeV is derived. This cross section limit can be
transformed into an upper limit for monopole pair production
of $\sigma_{IPC} \leq 10 \mu$b, assuming an IPC coefficient 
of $\eta =$ 0.03 for $^{181}$Ta and a
transition energy of 1.8 MeV. This limit is unfortunately
not sensitive enough to exclude E0 transitions from our measured 
\pn--sum-energy spectra definitely. 

The observed \pn{}--sum-energy line centered at an energy of 
$\sim$558 keV originates
actually from several close-lying transitions in the Ta-like nucleus
with energies between 1500 and 1580 keV, as suggested by the $\gamma$-ray 
spectrum shown in Fig. 8a.
Assuming that the remaining intensity excess of the observed $\sim$558 keV 
sum-energy line is due to several close-lying E0 transitions, the
corresponding K-conversion lines are expected to be spread out 
over a width of $\sim$80 keV, centered around 1470 keV 
with a total cross section of (0.7 $\pm$ 0.2)~mb. 
The cross section limit of 0.3 mb for Ta-like nuclei
and E$_{e^-} \approx$ 1.8 MeV given by 
Ditzel et al.~\cite{Dit96}, however, refers to a
line width of 20 keV. 
Even if one assumes that this cross section limit
is also valid for E$_{e^-} \approx$ 1500 keV, 
K-conversion lines with a total cross section of $\sim$0.7~mb,
spread out over an energy range of about 80 keV, could not 
be excluded by the measurements of Ditzel et al.~\cite{Dit96}. 

Another open problem remains by considering IPC transitions
in $^{238}$U, which has been used as a heavy-ion beam in all our previous
experiments
to study \pn{} pair creation in heavy collision systems.
In \UU
collisions an \pn--sum-energy line has been observed at an energy
of $\sim$815 keV~\cite{Ikoe93}.
For this observation two selection criteria 
were applied: the line was observed only at \pn{}
emission angles nearly $180^{\circ}$ and was enhanced by selecting  
negative energy differences. 
Both criteria resemble conspicuously those applied to the spectrum shown in 
Fig. 5b, and might
therefore indicate IPC transitions in $^{238}$U or in U-like nuclei. 
Unfortunately, this collision system was also investigated with the 
low resolution \mbox{Na(Tl)I} $\gamma$--ray detector, which does not allow
to reveal weak $\gamma$ lines in the measured spectra. 
On the other
hand we have not observed a corresponding 1835 keV line in the
$\gamma$-ray spectra after Doppler-shift correction to the U-like ions in
\UPb and \UTa collisions studied in this work. 
Consequently, only a
monopole transition in $^{238}$U could explain the previous observation
in \UU collisions provided that these lines were due to IPC transitions.
The observed 1780 keV $\gamma$ line could be
the $\gamma$ transition from the excited $0^+$ state to the
45 keV $2^+$ excited state of $^{238}$U adding up to a $0^+$
excitation energy of about 1825 keV, which could give rise to 
a 805 keV $0^+$ - $0^+$ monopole transition in close resemblance
to the previous observation of a \pn{} line at about 815 keV
in \UU collisions.

Finally it should be emphasized that the conjectured possibility of the
appearance of weak \pn--sum-energy lines due to discrete nuclear
transitions has been addressed for the first time in the ORANGE
experiments (see e.g. Ref. \cite{Ikoe93}).
Extensive Monte Carlo studies have revealed that IPC transitions
from a moving heavy nucleus can lead to narrow lines in the measured
\pn--sum-energy spectra, even without a Doppler-shift correction, when
restricting the \pn{} opening angles to the region around 180$^\circ$
(see Fig. 7 in Ref. \cite{Ikoe93}).
Our failure to identify such weak IPC transitions in the previous
experiments is not yet well understood.
It might be due  
to the limited experimental resolution
of the Doppler-shift technique, most noticeably in the determination 
of the 
azimuthal angles for the leptons and heavy ions
($\Delta\phi=60^{\circ}$ compared to $\Delta\phi=20^{\circ}$ 
in the new measurements), applied at that time to the
measured $\gamma$-ray and \pn--sum-energy spectra. 
For instance, in the case of our previous setup with an overall 
angular resolution of  
$\Delta\phi = 60^{\circ}$, the FWHM of the
Doppler-shift corrected \pn--lines is increased by about 10 keV.

In conclusion, the main progress of this work is the observation
of weak IPC \pn{} lines together with the accompanying $\gamma$ 
transitions.
Furthermore we may state that the weak \pn{} lines, reported in
previous experiments by the Orange collaboration~\cite{Ikoe93}, could 
have been 
caused most likely by weak IPC transitions, selected by the nearly 
Doppler-shift-free $180^{\circ}$ measurements of the Orange setup 
and the negative \pn{} energy differences. 
The only exception seems to be the observation
of a strong \pn{} line at $\sim$630 keV found previously in \UTa 
collisions~\cite{Ikoe93} which, however, could not be
reproduced by our improved experiments~\cite{Lein96b}.

\vspace*{1.5cm}
{\bf Acknowledgement:}
{\em We would like to thank all the people of the UNILAC accelerator
 operating crew  for their efforts in delivering stable $^{238}$U
 beams with high intensities.}

\newpage

\centerline{\Large \bf Figure Captions}                              

\vspace{1.5cm}
\noindent
{\bf Fig. 1.} Schematic view of the ORANGE 
spectrometers. The setup consists of
the following components: two iron--free orange-type 
magnetic spectrometers, 
two Si-(PIN)-diode arrays (PAGODAs) for lepton
detection, 19 PPACs to count the scattered heavy ions, 
an intrinsic Ge 
detector for $\gamma$-ray detection and a rotating 
target wheel
(see text for details).

\vspace{1.0cm}
\noindent
{\bf Fig. 2.} 
Experimental signature of the detected positrons. 
The abscissa shows the normalized difference between the 
momentum of the focused
positrons calculated from the pulse height of the Si 
detectors and their
momentum determined from the magnetic field of the 
spectrometer.
Only events falling in a narrow window centered around 
zero are 
accepted in the analysis.

\vspace{1.0cm}
\noindent
{\bf Fig. 3.} 
\pn--sum-energy spectra obtained by a Monte Carlo 
simulation of the 
1844 keV IPC (E1) transition in $^{206}$Pb
for \UPb collisions and peripheral ion kinematic
($\theta_{Pb,cm}=$50$^\circ$--70$^\circ$) in the laboratory 
(solid distribution)
and after an event-by-event Doppler-shift correction to 
the emitting
Pb nucleus (dashed distribution). The broad distribution 
in the laboratory
is centered around 815 keV and has a FWHM of $\sim$90 
keV, whereas the 
corrected one appears at $\sim$822 keV with a FWHM of 
$\sim$ 16 keV. 

\vspace{1.0cm}
\noindent
{\bf Fig. 4.} 
{\bf a)} Measured spectrum of $\gamma$ rays from \UPb  
collisions after 
Doppler-shift correction to the Pb-like ions, scattered 
in the R$_{min}$ 
range from 23 to 32 fm.
The line at (1844 $\pm$ 1) keV belongs
to the E1 transition 3$^-$ (2.65 MeV) 
$\rightarrow$ 2$^+$ (0.80 MeV) in $^{206}$Pb.\\ 
{\bf b)} The excitation probability of the 1844 keV 
$\gamma$ transition as a function of R$_{min}$.\\
{\bf c)} The Doppler-shift corrected \pn--sum-energy 
spectrum obtained 
under the assumption that Pb-like nuclei are the 
emitters. 
For R$_{min}=$ 23 --32 fm
and \pn{} energy differences from $-$200 to 175 keV. The 
contribution of
random coincidences ($\sim$ 15\%) is subtracted from the 
data.
The smooth solid line is a reference continuous 
distribution gained by event
mixing.

\vspace{1.0cm}
\noindent
{\bf Fig. 5.}
{\bf a)}
Uncorrected e$^+$e$^-$-sum-energy spectrum 
for the whole
opening angle region accepted 
($\theta_{\pn} = 40^{\circ} - 180^{\circ}$) 
taken with the collision system \UPb. 
The spectrum was obtained under identical conditions compared to 
that of Fig. 4c.\\ 
{\bf b)}
Uncorrected e$^+$e$^-$-sum-energy spectrum measured 
in \UPb collisions.
The full-line  histogram corresponds to the opening-angle bin of
$180^{\circ}$ ($\theta_{\pn} = 140^{\circ} - 180^{\circ}$), while 
the dashed-line histogram displays
the remaining opening-angle range of observation. The latter was
adjusted in the height to fit the total intensity of the spectrum obtained
for the $180^{\circ}$ bin.
The R$_{min}$ values range from 23 to 32 fm and $\Delta$E $\leq$ 175 keV. \\
{\bf c)} Uncorrected e$^+$e$^-$-sum-energy distribution for the 820 keV IPC 
transition in $^{206}$Pb obtained by a
Monte Carlo simulation for the opening angle range
$\theta_{\pn} = 140^{\circ} - 180^{\circ}$ and
R$_{min} =$ 23 $-$ 32 fm.

\vspace{1.0cm}
\noindent
{\bf Fig. 6.} {\bf a)} Doppler-shift corrected 
$\gamma$-ray spectrum measured 
for rather central collisions (R$_{min} \leq$ 20 fm). 
The line at (1770 $\pm$ 1) keV corresponds to the M1 
transition 
$\frac{7^-}{2}$ (2.34 MeV) $\rightarrow$
$\frac{5^-}{2}$ (0.57 MeV) in $^{207}$Pb, produced 
 by 1n-transfer reaction.\\
{\bf b)} Excitation probability as a function of 
R$_{min}$ of the 1770 keV
$\gamma$ transition. \\
{\bf c)} The same as in Fig. 4c, but for R$_{min}$ 
values between 17 and 20 fm.

\vspace{1.0cm}
\noindent
{\bf Fig. 7.} {\bf a)} $\gamma$--ray spectrum from the collision 
system \UPb after Doppler-shift correction to the U-like ion, scattered
in the R$_{min}$ range 19.5--23 fm. 
A line appears at
an energy of (1778 $\pm$ 2) keV.\\
{\bf b)} Excitation probability as a function of R$_{min}$ of
the $\gamma$ line at 1778 keV.\\
{\bf c)} The Doppler-shift corrected \pn--sum-energy 
spectrum obtained 
by assuming that U-like nuclei are the 
emitters 
for R$_{min}=$ 19.5--23 fm
and \pn{} energy differences from $-$200 to 175 keV.

\vspace{1.0cm}
\noindent
{\bf Fig. 8.} {\bf a)} Doppler-shift corrected $\gamma$--ray spectrum 
observed in the collision system \UTa.  
Ta-like recoiling ions for rather peripheral collisions
(R$_{min} =$ 22.4--30 fm) are assumed to be the emitter. 
Several lines appear at energies
below 1600 keV.\\
{\bf b)} Excitation probability as a function of
R$_{min}$ of the line at 1380 keV.

\vspace{1.0cm}
\noindent
{\bf Fig. 9.} {\bf a)} \pn{}--sum-energy spectrum obtained after Doppler-shift
correction to the Ta-like ion in the R$_{min}$ range
22.4--30 fm and \pn{} energy differences from
$-$150 to +100 keV. The spectrum was measured in coincidence
with only one scattered heavy ion.\\ 
{\bf b)} Doppler-shift corrected
\pn--sum-energy spectra obtained by a Monte Carlo
simulation of the transition in the Ta-like nucleus centered around 560 keV.
The underlying kinematical conditions are identical to that of Fig. 9a.
The rather narrow line width of 30--35 keV is caused by the
spectrometer momentum acceptance. 

\vspace{1.0cm}
\noindent
{\bf Fig. 10.} Uncorrected \pn{}--sum-energy spectrum measured with the
collision system \UTa. The full-line histogram corresponds to the 
opening-angle bin of $180^{\circ}$. The dashed-line histogram displays the 
remaining opening-angle range of observation. The latter was adjusted
in the height to fit the total intensity of the spectrum obtained
for the $180^{\circ}$ bin.
The R$_{min}$ values range from 22.4 to 30 fm and the energy differences
are restricted to  $-$150 keV $\leq \Delta$E $\leq$ +100 keV.

\newpage
\pagestyle{empty}

\vspace*{4.5cm}
\begin{center}
 \epsfig{file=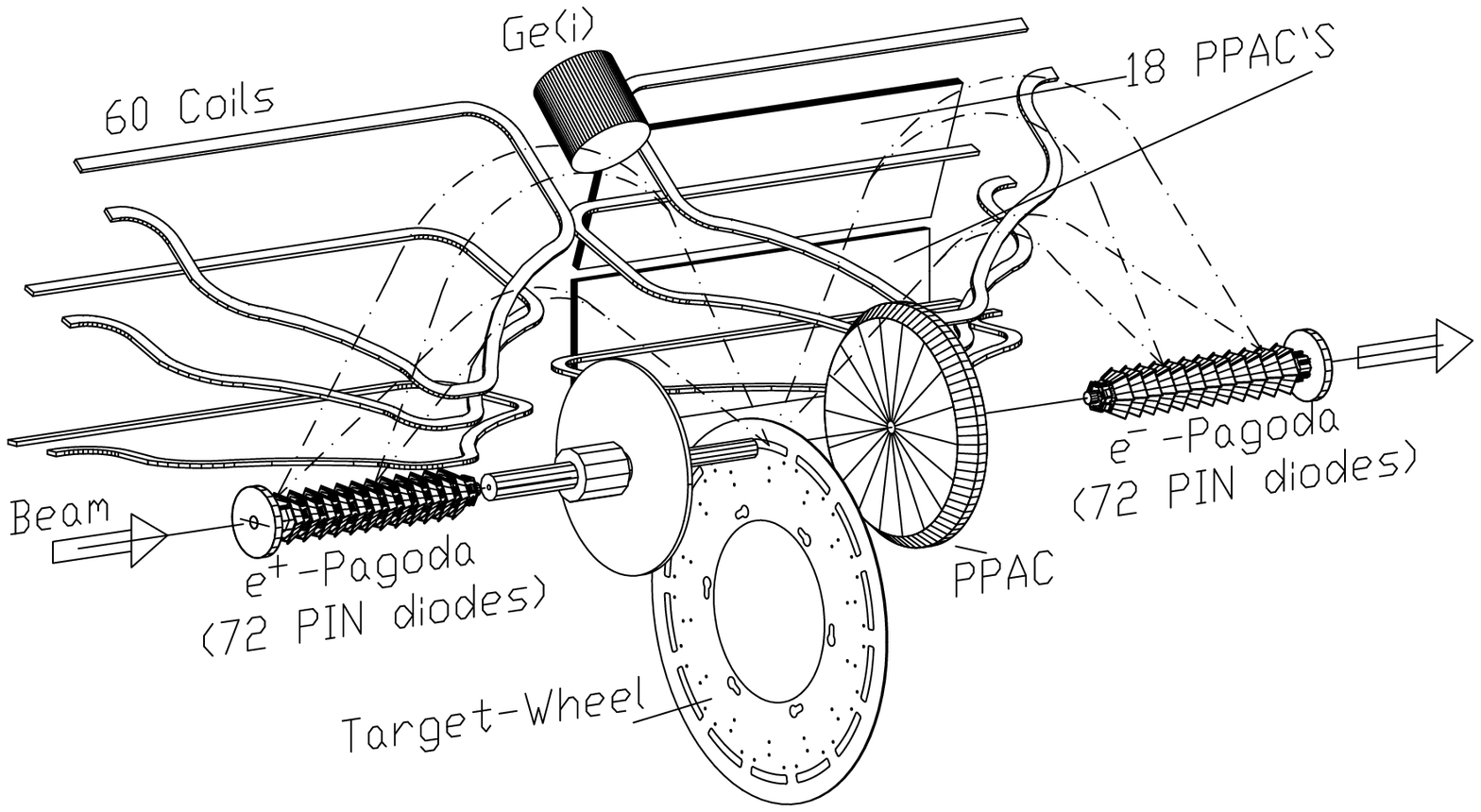,width=1.05\linewidth}
\end{center}

\vspace*{5.0cm}
{\Large \bf Figure 1}

\vspace*{3 mm}
(S. Heinz {\em et al.}, Z. Physik A)

\newpage 
\pagestyle{empty}

\vspace*{7cm}
\begin{center}
 \epsfig{file=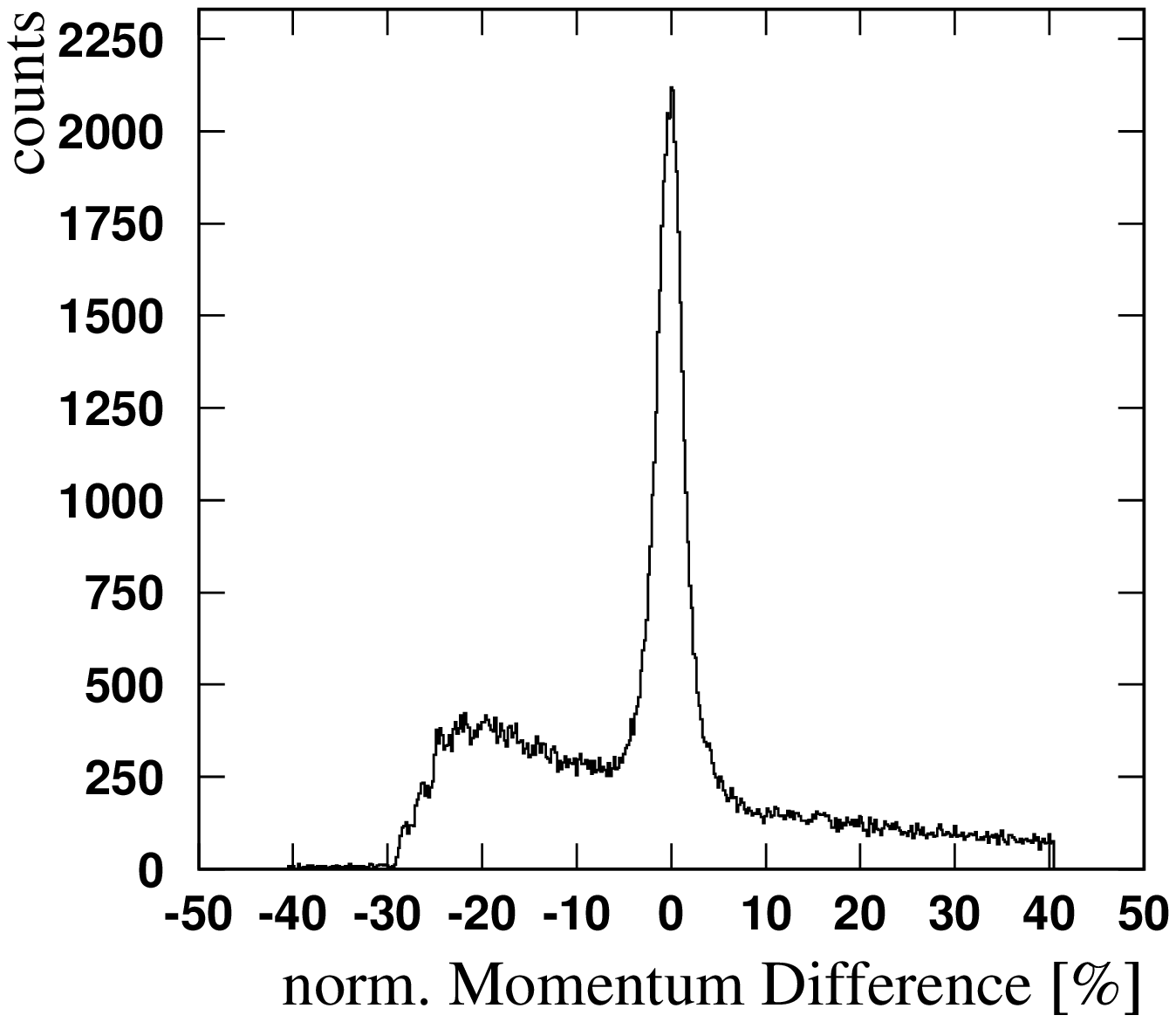,width=11 cm}
\end{center}

\vspace*{0.5cm}
{\Large \bf Figure 2}

\vspace*{3 mm}
(S. Heinz {\em et al.}, Z. Physik A)

\newpage 
\pagestyle{empty}

\vspace*{4cm}
\begin{center}
 \epsfig{file=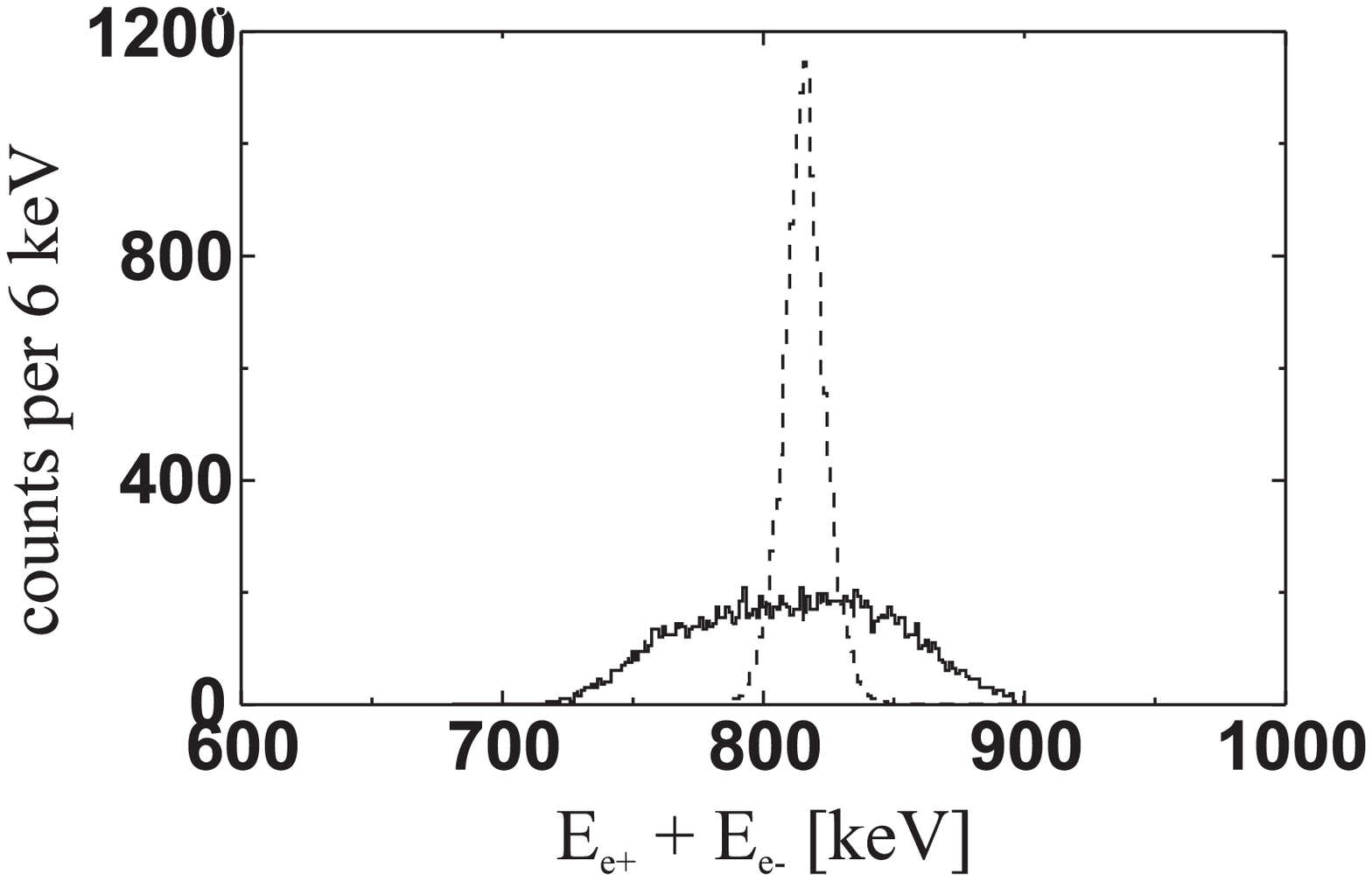,width=11 cm}
\end{center}

\vspace*{0.5cm}
{\Large \bf Figure 3}

\vspace*{3 mm}
(S. Heinz {\em et al.}, Z. Physik A)

\newpage 
\pagestyle{empty}

\begin{center}
 \epsfig{file=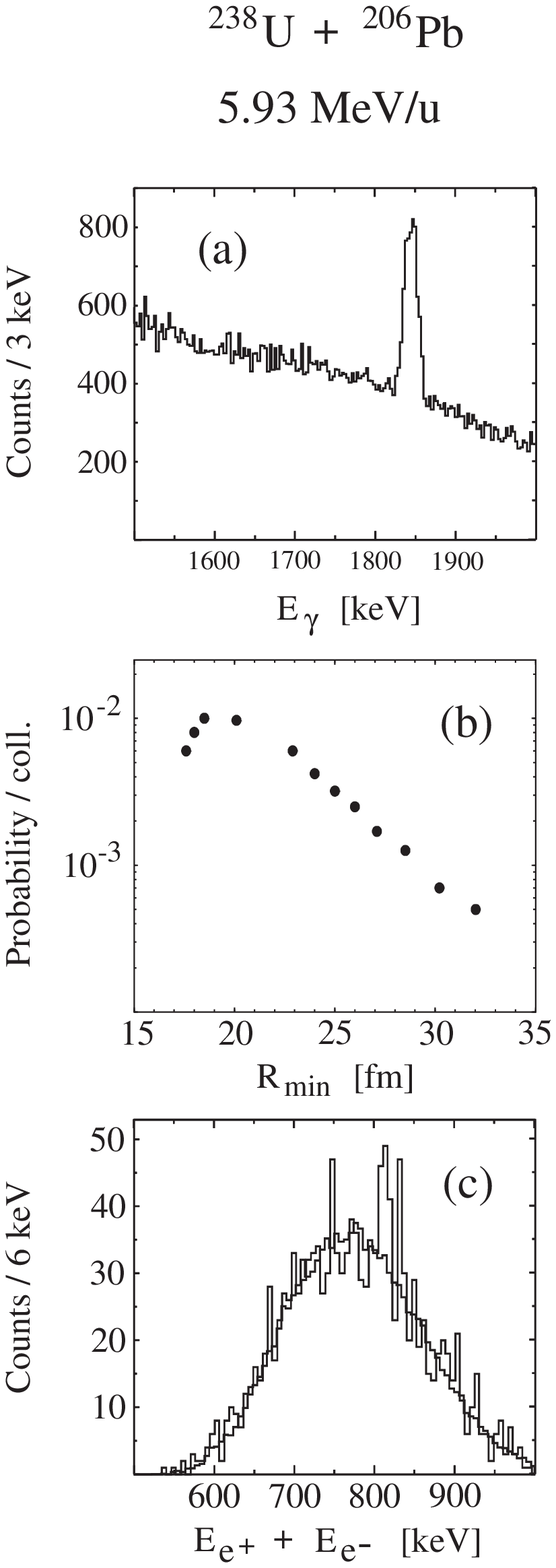,height=21 cm}
\end{center}

\vspace*{0.3cm}
{\Large \bf Figure 4}

\vspace*{3 mm}
(S. Heinz {\em et al.}, Z. Physik A)

\newpage
\pagestyle{empty}

\begin{center}
 \epsfig{file=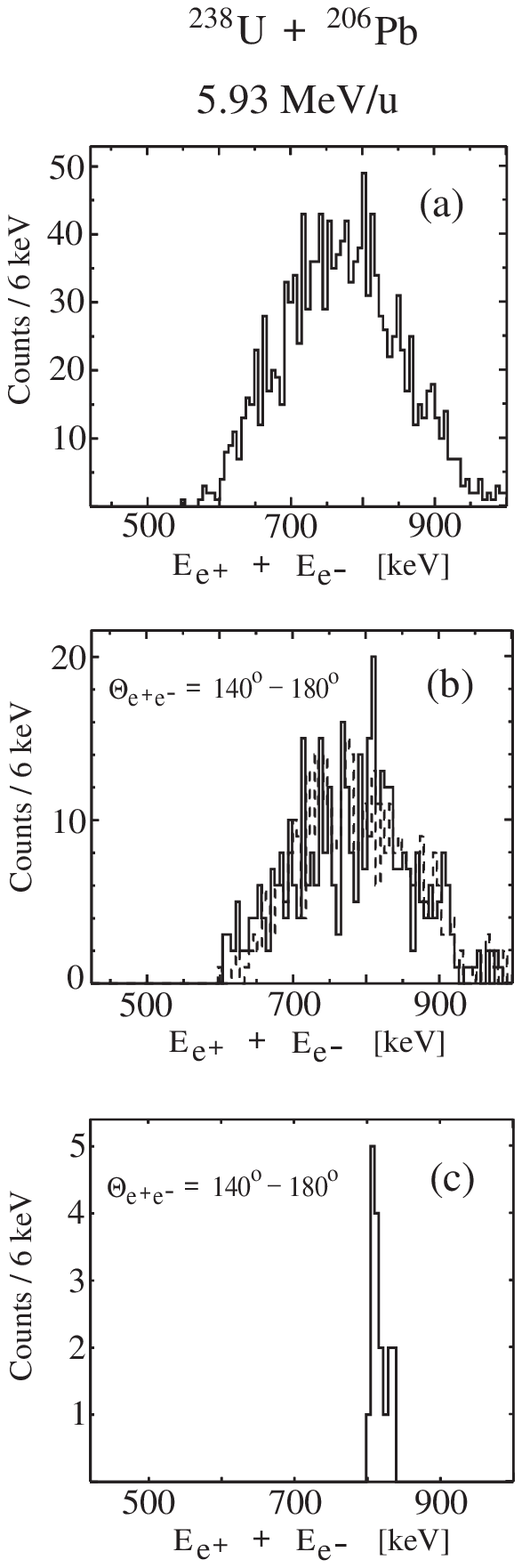,height=21 cm}
\end{center}

\vspace*{0.3cm}
{\Large \bf Figure 5}

\vspace*{3 mm}
(S. Heinz {\em et al.}, Z. Physik A)
\newpage
\pagestyle{empty}

\begin{center}
 \epsfig{file=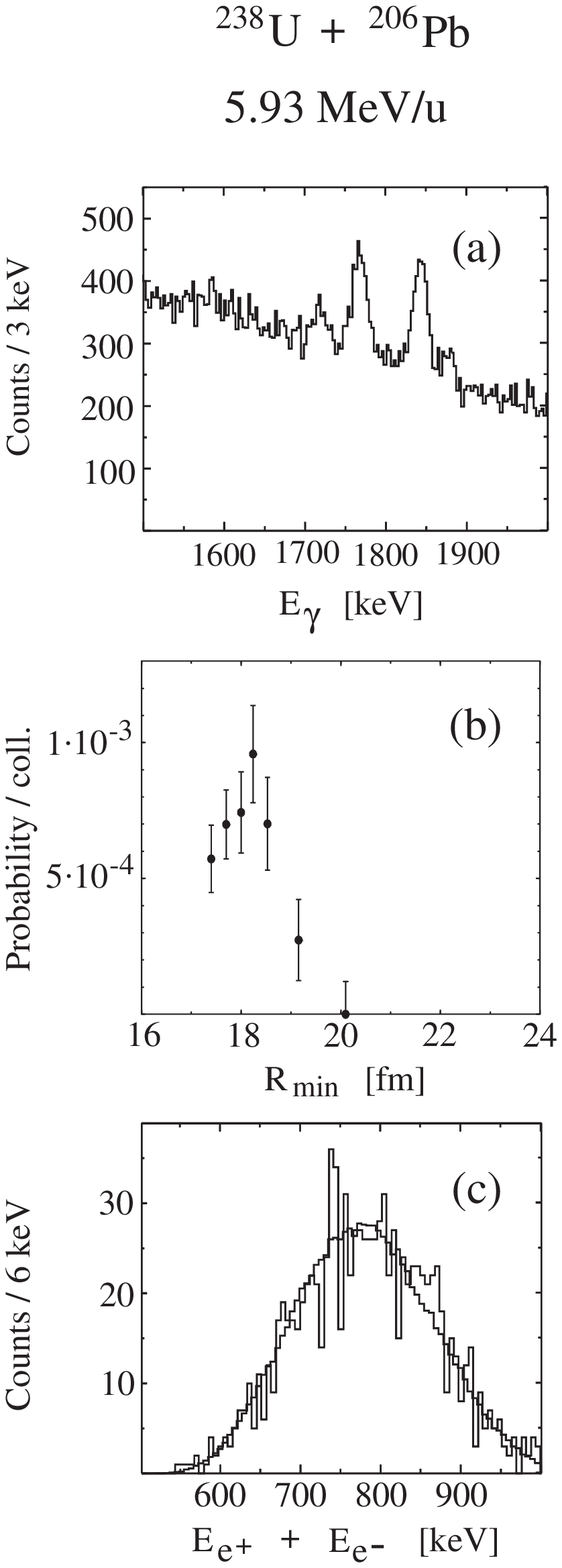,height=21 cm}
\end{center}

\vspace*{0.3cm}
{\Large \bf Figure 6}

\vspace*{3 mm}
(S. Heinz {\em et al.}, Z. Physik A)

\newpage
\pagestyle{empty}

\begin{center}
 \epsfig{file=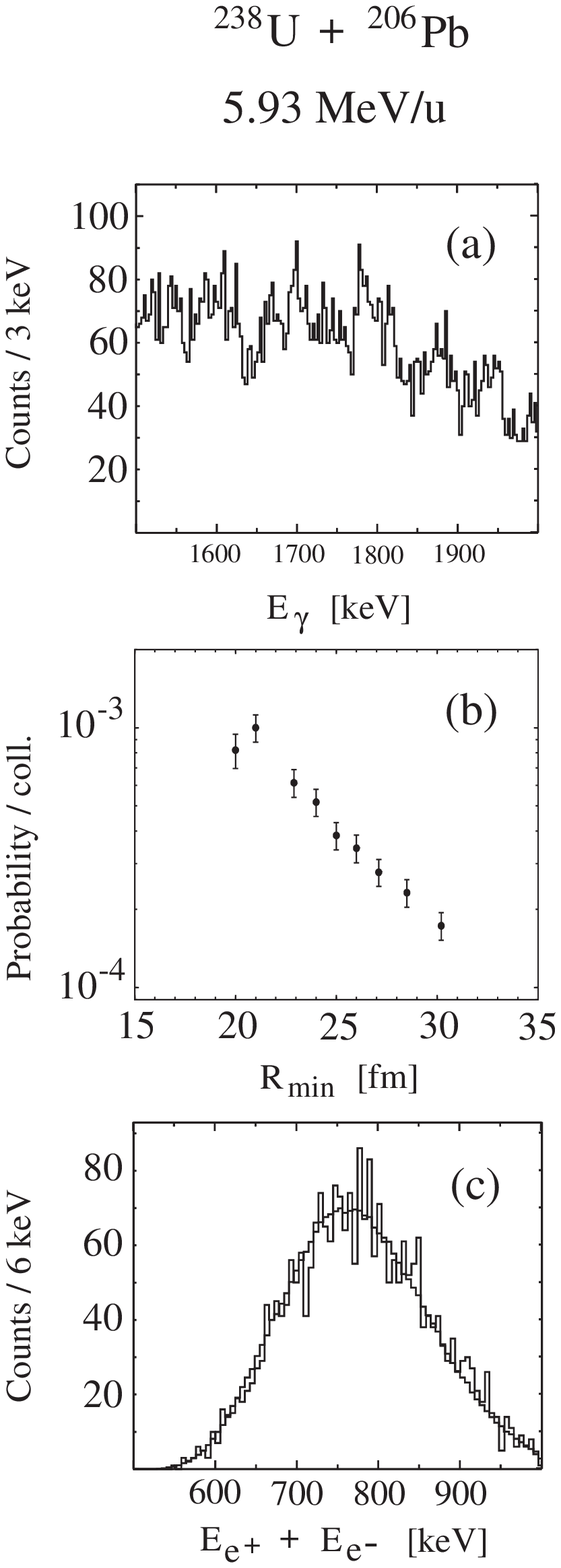,height=21 cm}
\end{center}

\vspace*{0.3cm}
{\Large \bf Figure 7}

\vspace*{3 mm}
(S. Heinz {\em et al.}, Z. Physik A)
\newpage
\pagestyle{empty}

\begin{center}
 \epsfig{file=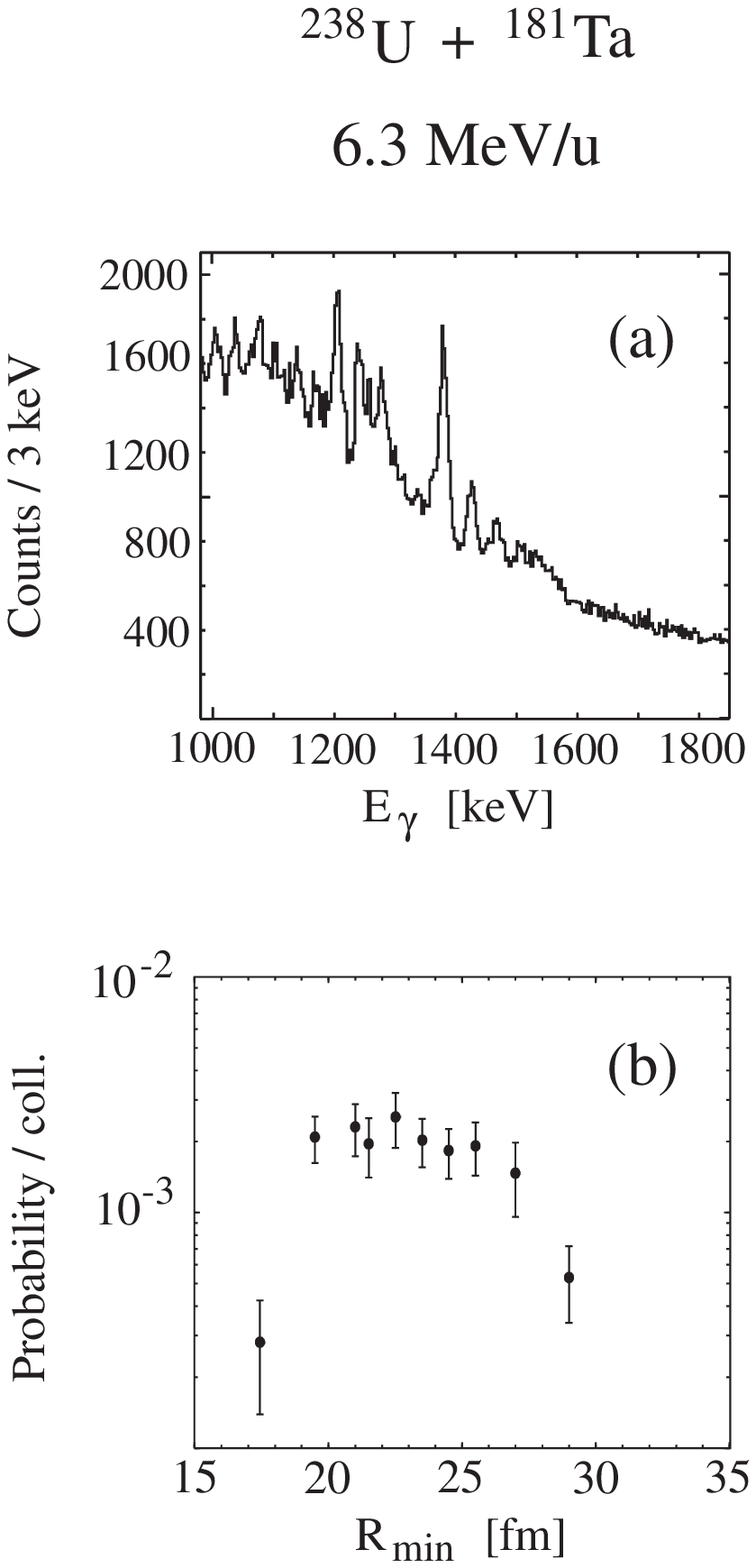,height=21 cm}
\end{center}

\vspace*{0.3cm}
{\Large \bf Figure 8}

\vspace*{3 mm}
(S. Heinz {\em et al.}, Z. Physik A)

\newpage
\pagestyle{empty}

\begin{center}
 \epsfig{file=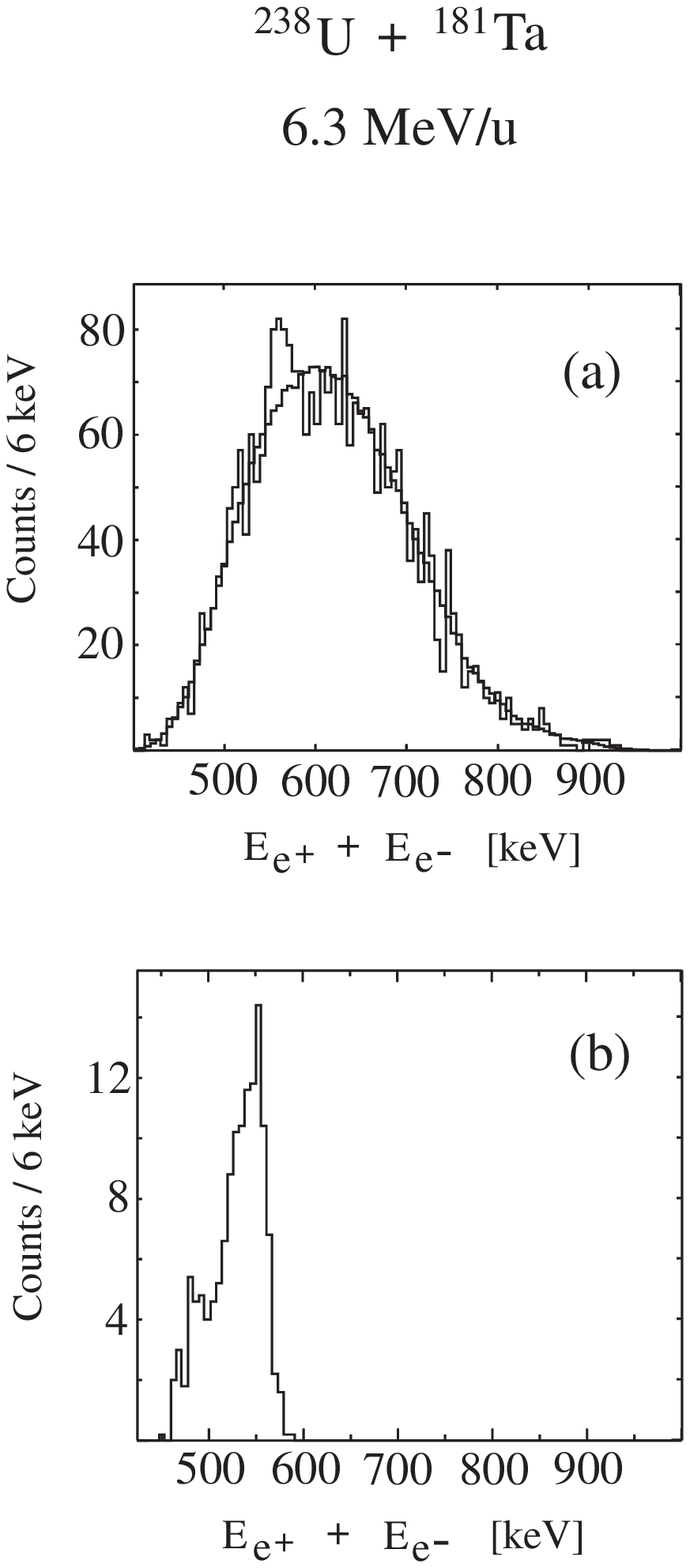}
\end{center}

\vspace*{0.3cm}
{\Large \bf Figure 9}

\vspace*{3 mm}
(S. Heinz {\em et al.}, Z. Physik A)

\newpage
\pagestyle{empty}

\begin{center}
 \epsfig{file=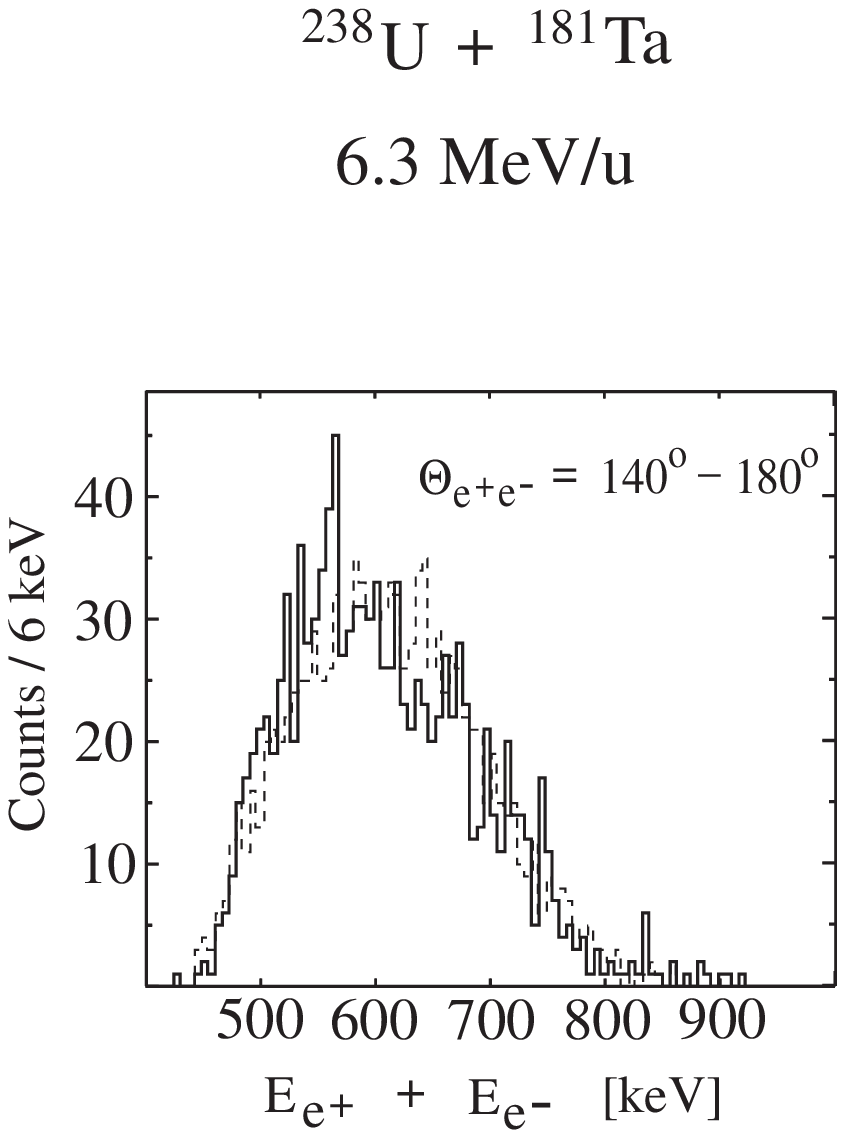}
\end{center}

\vspace*{0.3cm}
{\Large \bf Figure 10}

\vspace*{3 mm}
(S. Heinz {\em et al.}, Z. Physik A)

\end{document}